\documentclass[preprint]{sigplanconf}
\usepackage{amsmath}
\usepackage{amsfonts}
\usepackage{hyperref}
\usepackage{booktabs} % For formal tables
\usepackage{balance}
\usepackage{multirow}
\usepackage{graphicx}
\usepackage{float}
\usepackage{enumitem}
\usepackage{color}

\newcommand{\sparagraph}[2][.]{\vspace{1mm}\noindent {\bf #2#1}}

\newcommand{\bad}[1]{\textcolor[RGB]{139,0,0}{#1}}

\newcommand{\good}[1]{\textcolor[RGB]{0,139,0}{#1}}

\begin{document}

\setlength{\pdfpageheight}{\paperheight}
\setlength{\pdfpagewidth}{\paperwidth}

\setcopyright{acmlicensed}  % default

\title{Learned Garbage Collection}

\authorinfo{Lujing Cen$^1$, Ryan Marcus$^{12}$, Hongzi Mao$^1$, \\ Justin Gottschlich$^2$, Mohammad Alizadeh$^1$, Tim Kraska$^1$}
{$^1$MIT CSAIL \quad $^2$Intel Labs}
{\{lujing, ryanmarcus, hongzi, alizadeh, kraska\}@csail.mit.edu \\ justin.gottschlich@intel.com}
\maketitle

\begin{abstract}
Several programming languages use garbage collectors (GCs) to automatically manage memory for the programmer. Such collectors must decide when to look for unreachable objects to free, which can have a large performance impact on some applications.
% While expert users can manually tune GCs, the process is arduous and inaccessible to many users.
In this preliminary work, we propose a design for a learned garbage collector that autonomously learns over time when to perform collections. By using reinforcement learning, our design can incorporate user-defined reward functions, allowing an autonomous garbage collector to learn to optimize the exact metric the user desires (e.g., request latency or queries per second). We conduct an initial experimental study on a prototype, demonstrating that an approach based on tabular Q learning may be promising.
\end{abstract}

% TODO: Add keywords.
%\keywords{code similarity, program synthesis, machine programming, software development, software maintenance}

\section{Introduction}
In many programming languages, automatic memory management is performed via \emph{garbage collection} (GC). \emph{Generational GC}, in which short-lived objects are stored in early generations and longer-lived objects are slowly moved into later generations~\cite{gc_gen}, is a common technique used in languages like Java and Python. A generational GC's \emph{collection policy} -- when the GC chooses to scan the heap for objects that can be discarded -- vary in complexity. For example, Java's garbage collector is sophisticated, representing significant engineering effort~\cite{java_gc}. On the other hand, Python's garbage collector is simpler, using a combination of reference counting and generational "stop-the-world cycle detection" at predefined intervals with three generations~\cite{mark_and_sweep}.

For some programs, the location where GC is performed can impact performance~\cite{gc_search_opt}. For example, if a function creates many allocations that will soon have their reference count reach zero (triggering an automatic collection), garbage collection may waste time scanning these allocations. Alternatively, consider a web server: performing GC in the middle of a request may deteriorate response latency~\cite{gc_server}.

Sophisticated users running large-scale CPython applications will often tune or modify CPython's GC policy to suit their performance requirements~\cite{url-instagram-gc}, such as tail request latency. Yet, many users may not even know the garbage collector exists, or how to effectively tune it. Here, we ask: \emph{can efficient custom-tailored generational GC policies be discovered automatically, with minimal human interaction?} Such an automatic system could improve the performance of user applications, and could potentially lessen the engineering burden required to implement GC in new languages.

We present a preliminary sketch and prototype of a garbage collection policy powered by reinforcement learning (RL). 
While learning approaches to GC are not new~\cite{gc_rf, gc_search_opt}, existing approaches require pre-training on captured execution traces, and try to minimize the time spent in GC mechanisms. We target long-running programs that repeatedly execute a core loop, such as a web server or a database, and we attempt to optimize a custom reward function, such as request latency or video delay. This potentially enables the automatic learning of custom-tuned GCs without human intervention. For example, in a soft real time system, such as a video game, our system could learn to perform GC outside of the critical loop, which may require more GC time, but may have a smaller impact on the user’s experience. Surprisingly, we found that classical tabular Q learning~\cite{q} is sufficient for quickly learning advanced GC policies in our test applications, including several toy programs and a real-world project management toolkit.

In this paper, we highlight some of the technical concerns and early solutions we discovered when applying reinforcement learning to garbage collection. Specifically, we found that correctly tuning the algorithm’s prior beliefs (e.g., biasing the system towards one action or another), along with a small amount of reward shaping (e.g., modifying the feedback signal to induce good behavior), are critical to fast convergence. The contributions of this paper are:

\begin{itemize}
    \item{To the best of our knowledge, we present the first end-to-end learned garbage collection policy trained entirely with reinforcement learning, while the program is executing. Our prototype attempts to optimize a user-provided and application-specific reward function, customizing a GC policy for the user's domain.}
    \item{We highlight how a modified prior distribution and reward shaping can aid an RL-powered garbage collector in finding a good policy quickly.}
    \item{We present early experimental evidence suggesting that a learned GC policy can substantially outperform the (simple) default GC policy included with CPython on a variety of benchmarks.}
\end{itemize}

The rest of this paper is organized as follows: in Section~\ref{sec:formulation}, we present our problem formulation and precisely define the underlying Markov decision process (MDP). In Section~\ref{sec:gc}, we introduce our prototype garbage collector. We present early experimental results in Section~\ref{sec:experiments}, and provide concluding remarks and directions for future work in Section~\ref{sec:conclusions}.

%%% Local Variables:
%%% mode: latex
%%% TeX-master: "main.tex"
%%% End:

\section{Related Work}
Prior work by Jacek et al. gave a dynamic programming algorithm to compute the optimal GC locations (in terms of reducing GC execution time) given an execution trace of a program on a particular input~\cite{gc_search_opt}. Later, Jacek et al. showed that supervised learning techniques could be used to generalize these execution traces to other inputs~\cite{gc_rf}. While these techniques can be helpful foived programs for which trace information is available,  both works (1) require post-hoc analysis of program traces, and do not automatically adapt their policies in real time, and (2) minimize time spent in GC mechanisms, as opposed to optimizing a user-defined reward function (e.g., request latency).

Many previous works have applied reinforcement learning to various systems problems, including query optimization~\cite{neo,qo_state_rep}, cluster scheduling~\cite{decima}, stream processing~\cite{lift}, and cloud provisioning~\cite{perfenforce_demo,wisedb-cidr}. More broadly, applying reinforcement learning to systems problems can be viewed as machine programming~\cite{pillars}, autonomously inventing new policies and adapting to new environments.

Garbage collection is an active area of research~\cite{java_gc, java_gc2, mark_and_sweep, gc_v8}. \cite{gc_api} proposes new user-facing APIs to more easily (manually) customize garbage collectors. \cite{gc_gpu} suggests offloading mark-and-sweep style computations to a GPU. 
 
%%% Local Variables:
%%% mode: latex
%%% TeX-master: "main"
%%% End:

\section{Overview \& Formulation}
\label{sec:formulation}
In this section, we explain a simple model of a generational garbage collector. Then, we give an overview of our approach and specify the underlying Markov decision process (MDP).

\begin{figure}
    \centering
    \includegraphics[width=\linewidth]{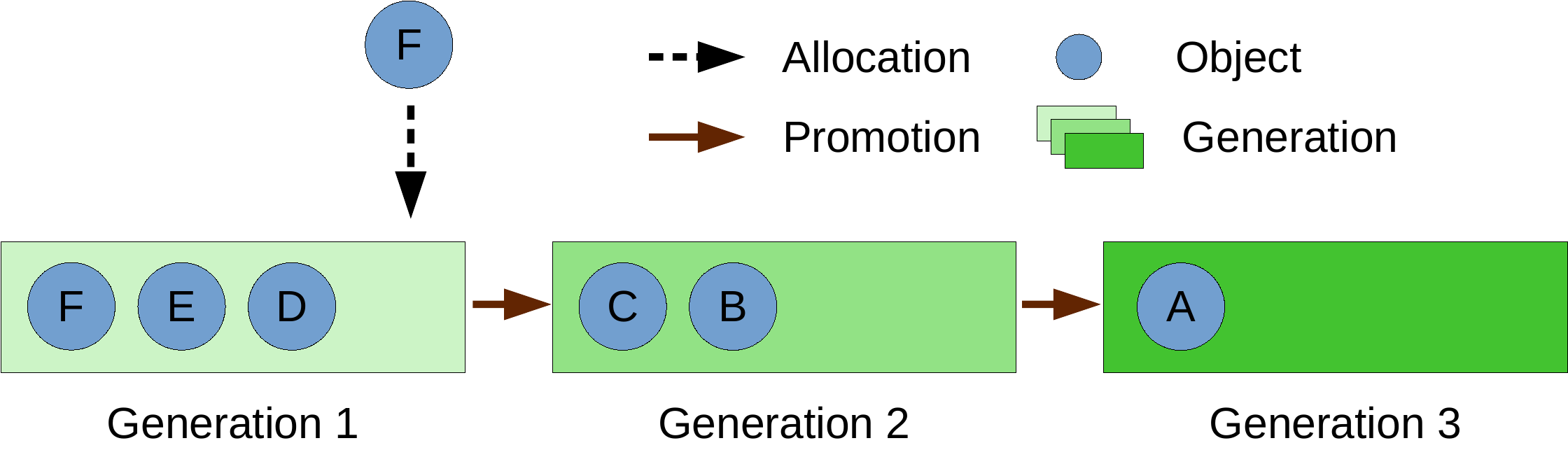}
    \caption{Generational garbage collection. New objects are allocated in generation 1. When a generation is collected, mark-and-sweep determines which allocations are freed and which ``survive.'' Surviving allocations are promoted to the next generation, ending with generation 3.}
    \label{fig:gen_gc}
\end{figure}

\sparagraph{Generational garbage collectors}
Our model of a generational garbage collector is shown in Figure~\ref{fig:gen_gc}. When a new object is allocated (F), it is placed in generation 1.  Before allocation occurs, a \emph{garbage collection policy} determines whether or not a particular generation is collected.\footnote{A policy may decide to collect no generations at all.} We define \emph{collecting a generation} to mean: (1) determining which objects in that generation and all younger generations are no longer reachable (e.g., mark-and-sweep), (2) freeing those unreachable objects, and then (3) \emph{promoting} objects in that generation and all younger generations to the next generation (except for objects in generation 3, which remain until freed). 

Our model is agnostic to (1) whether or not orthogonal garbage collection methods are also used (e.g., reference counting), (2) whether entire allocations or just pointers are stored in the generational heaps (and thus whether pointers or entire objects are copied during promotion), and (3) the method of determining whether an object is reachable (e.g., parallel mark-and-sweep or pointer tracking). Our model does assume that the number of generations is constant and finite. We refer to each generation as $G_1, G_2, \dots, G_{|G|}$.

The goal of a garbage collection policy is to maximize a user-specified reward function, $R$. We assume that $R$ is moderately expensive to evaluate and noisy (e.g., can only be called every few seconds and may exhibit fluctuations). Without loss of generality, we assume that the range of $R()$ is $[0, 1]$. Example reward functions could be transactions-per-second or negative tail request latency. While maximizing $R()$ likely involves minimizing time spent in garbage collection, this may not always be the case: there may be opportune times to perform garbage collection that take slightly longer to collect but improve the user's reward function, such as immediately after a request is served or a file is updated. 

\begin{figure}
    \centering
    \includegraphics[width=\linewidth]{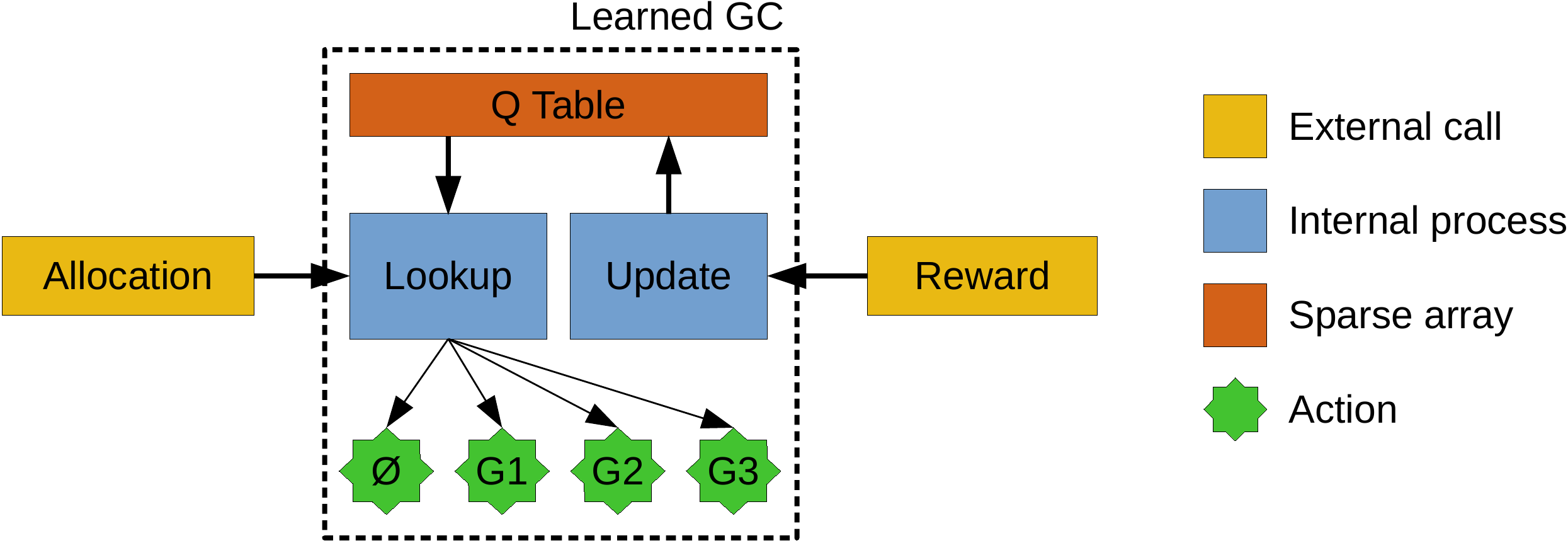}
    \caption{Learned GC architecture. When an allocation occurs, we lookup the appropriate entry in the Q table to select an action. Periodically, a user-provided reward signal is used to update the values in the Q table. }
    \label{fig:learned_gc}
\end{figure}

\sparagraph{Learned GC overview} Our learned garbage collector works via a feedback loop that enables continuous improvement. Figure~\ref{fig:learned_gc} shows an overview of how the collector works. 

When an \emph{allocation} occurs, our system performs one of two actions: \emph{(i)} collect nothing or \emph{(ii)} collect a particular generation. This decision is made via looking up a value in the \emph{Q table}, a multi-dimensional sparse array which stores the expected value\footnote{A balance of long-term and short-term rewards, detailed in Section~\ref{sec:gc}.} of each potential action, indexed by allocation sites and current memory usage. Intuitively, for a specific allocation site (e.g., line of code), memory usage (e.g., 50MB), and action (e.g., collect nothing), the Q table contains one value, representing the expected reward from performing that action in that situation. Actions are selected to maximize the expected reward.

Periodically, the user supplies the garbage collector with a \emph{reward}, a quantity the user wishes to maximize. For example, for a web application, the user may report the (inverse of) average request latency every 5 seconds. When reported, the learned GC \emph{updates} the Q table, propagating information about the user-provided reward signal to the appropriate cells of the Q table. If the reward improves, the Q table will be adjusted to favor the current policy. If the reward deteriorated, the Q table may be adjusted to explore other options.

\sparagraph{Markov decision process (MDP)} Here, we formulate our problem as an MDP, the classical formulation used by most reinforcement learning algorithms. For an overview of MDPs, see~\cite{rl_book}. Roughly, MDPs are composed of a set of states $S$ and a set of actions $A$. An agent begins in some initial state $s_0 \in S$. The agent may then choose an action $a_0 \in A$. Afterwards, the agent receives a reward $r_0$, and a new state, $s_1$. The agent's goal is to maximize the sum of its rewards over time, that is, at time $t$, the agent wishes to maximize $\sum_0^{t-1} r_t$.

In our context, each state of an MDP contains features describing the current allocation. In general, states may include any amount of information, such as the allocation site (e.g., program counter position), current memory usage, the time since the last collection, the size of each generation, etc. In this preliminary work, we define each state $s_i \in S$ to be a tuple of $s_i = (l_i, m_i)$ where $l_i$ is the location of a particular allocation (in terms of the program counter) and $m_i$ is the current memory usage of the program. Actions correspond to collecting a particular generation, or doing nothing: 

\begin{equation*}
A = \{CG_1, CG_2, \dots, CG_{|G|}, \emptyset\}
\end{equation*}

\noindent where $CG_n$ corresponds to collecting generation $n$.

Because the optimal GC policy for some programs may be to simply never collect, and to let memory grow unbounded, we also assume that there is a user-specified memory threshold $M$. If, during any allocation, the total memory usage of the program exceeds $M$, a collection of the oldest generation (a full collection) is forced, and the agent is penalized.

%%% Local Variables:
%%% mode: latex
%%% TeX-master: "main"
%%% End:

\section{Learned GC}
\label{sec:gc}

In this section, we describe how we apply and slightly modify classical tabular Q learning~\cite{q} to the MDP described in Section~\ref{sec:formulation}. We note that tabular Q learning is distinct from deep Q learning~\cite{dqn}, and does not involve a neural network. We first explain the layout of the Q table, then how inference (decision-making) is performed using the table, then how the table is updated based on the user-defined reward function.

\sparagraph{Q table} The Q table is an array mapping every possible state and action combination to a real number, $Q: S \times A \to \mathbb{R}$, representing a belief that a particular action should be chosen in that state (higher values represent a stronger belief). While the size of the state space is large (i.e., every possible allocation site and every possible amount of memory usage), we note that storing $Q$ in memory is often tractable. First, while a program may contain millions of possible allocation sites (e.g., lines of code or program counter locations), the number of those sites where an allocation actually occurs is smaller. Second, we discretize the memory usage into a fixed number of bins. The result creates a Q table that is of manageable size even for large applications (e.g., under 16MB for a large, production-scale application). Each value in the Q table is initially set to zero.

\sparagraph{Inference} When an allocation occurs at state $s_i = (l_i, m_i)$, with location $l_i$ and current memory usage $m_i$ (discretized), the optimal action according to the current Q table is:
\begin{equation*}
Opt(s_i) = \max_{a \in A} Q(s_i, a)
\end{equation*}

Choosing to perform the optimal action $Opt(s_i)$ for every allocation represents a strategy of pure \emph{exploitation}. This amounts to assuming that the Q table contains the correct values in each cell. In stochastic environments, such a strategy is likely to "get stuck" in a local minima: that is, find some policy that works better than any small perturbation to that policy, but is still not the ideal policy. To avoid this, we use an \emph{epsilon-greedy} policy~\cite{rl_book} to encourage \emph{exploration}. An epsilon-greedy policy performs a random action with probability $\epsilon$, and plays according to the Q table with probability $1 - \epsilon$. Over time, as the Q table's cells are populated by better and better data, the value of $\epsilon$ is decayed.

\sparagraph{Updates} When a reward is received from the user-defined reward function, the cells of the Q table are updated. Since allocations are performed more frequently than the reward is measured, we must \emph{attribute} parts of the reward to each action. To do this, we naively assign the entire reward  to each action taken since the last time the reward was measured. Then, for each state $s_i$ and each action $a_i$ executed,  we update each cell in the Q table using the classical update rule:
\begin{equation*}
    Q(s_i, a_i) = Q(s_i, a_i) + \alpha \times \left( r + \gamma \times Opt(s_{i+1}) - Q(s_i, a_i) \right)
\end{equation*}

\noindent where $0 < \alpha \leq 1$ is the \emph{learning rate} and $0 < \gamma \leq 1$ is the \emph{discount factor}. The learning rate $\alpha$ controls how much old information is valued against new information (larger values put more weight on new information). In the context of garbage collection for long-running services, the ideal policy is unlikely to shift quickly in a short period of time, so a small learning rate is likely appropriate. The discount factor $\gamma$ controls how  short-term rewards are valued against long-term rewards. Because of our simple attribution scheme, and because we generally care about the long-term performance of long-running programs, a large $\gamma$ is likely appropriate.

\subsection{Optimizations \& Discussion}
\label{sec:optimizations}
Using the unmodified Q learning algorithm, as described above, can lead to initially catastrophic policies. Since the agent initially chooses actions entirely at random, the agent's initial performance may be much worse than a naive policy. However, we have prior knowledge about how a garbage collector should behave. We next discuss three optimizations which integrate parts of the community's collective wisdom about garbage collection into our learned GC.

\sparagraph{Optimization \#1: priors (P)} For the majority of allocations, the correct action is not to collect any generation. In fact, by default, CPython collects at most every 700 allocations~\footnote{\url{https://docs.python.org/3/library/gc.html}}. Without any tuning, the epsilon-greedy policy described above will choose to collect some generation $\frac{3}{4}$ of the time -- a vast difference from CPython's default $\frac{1}{700}$. Thus, we propose a simple modification to the epsilon-greedy scheme to assign  more weight to the "collect nothing" action. 

\sparagraph{Optimization \#2: reward shaping (S)} Reinforcement learning algorithms can often benefit from small modifications to the reward function that do not change the optimal policy, but provide additional feedback to the agent~\cite{reward_shaping}. When the learned agent chooses to collect, we penalize the reward value slightly based on how long the collection process took. While minimizing time spent performing GC is not an explicit goal, doing so likely correlates with improvements to the user's reward signal (e.g., less time in GC means more time processing requests). Thus, if the agent, based on the rewards seen so far, is unsure whether one allocation site or another is better for performing a collection, this reward shaping encourages the agent to pick the allocation site that resulted in a lower GC time.

% For example, if the memory threshold $M$ is 200MB, then we know \emph{a priori} that choosing any action except a fully collection when the program's memory usage is at 200MB will result in poor expected value. We (lazily) initialize all cells in the Q table that represent $A \ {CG_{|G|}}$ at or above the memory threshold to -100, ensuring they are unlikely to be chosen.
\sparagraph{Optimization \#3: table initialization (I)} When a program's memory usage surpasses a user-defined threshold $M$, a collection is forced and the agent is penalized. Until the agent hits this penalty, the "never collect" policy  appears good; the agent only learns that the "never collect" policy is bad after encountering the penalty. However, we can "pre-teach" the agent about this penalty via specially initializing certain cells of the Q table. 
For example, if the memory threshold $M$ is 200MB and the program's current memory usage is 199MB, then we know \emph{a priori} that  choosing the $\emptyset$ "do nothing" action is likely to have a poor expected value. We (lazily) initialize all cells in the Q table that represent anything but a full collection ($CG_{|G|}$) when memory usage is at or above $M$ to -100, making them unlikely to be chosen.

%\subsection{Discussion}

\sparagraph[?]{Will this work for all programs} Garbage collectors are general-purpose, and must work with a wide variety of applications. Many reinforcement learning approach, especially the simple approach described here, requires seeing the same or similar states multiple times in order to explore different policies. This happens in long-running programs that execute a core loop, like a web server or a database. However, in short-lived programs, there may not be sufficient time or repetition to learn a policy. In future work, we plan on investigating learning between runs of the same program, or trying to learn a more general, program-agnostic policy.

\sparagraph[?]{Why not use neural networks} Modern Q learning techniques often use a neural network (NN), or other function approximator, instead of storing a table~\cite{dqn}. Using an NN has significant advantages when the state space is extremely large, and when the state is represented by a semantically-rich feature vector. We initially experimented with neural networks, but discovered that, in our case, the state space (containing an allocation site and current memory usage) is not unreasonably large, nor is it particularly semantically-rich (two allocation sites being near each other does not mean they behave similarly). Additionally, the tabular representation of the Q table allows for fast inference times: an action can be selected with only four array lookups and a max reduction, along with handling an epsilon-greedy policy. When NNs are used, this inference time is more expensive. Since allocations happen frequently (often more than hundreds per second), fast inference time is essential for GC.

\begin{equation*}
\end{equation*}

%%% Local Variables:
%%% mode: latex
%%% TeX-master: "main"
%%% End:

\vspace{-10mm}
\section{Experiments}
\label{sec:experiments}

\begin{figure*}
  \centering
  \includegraphics[width=0.97\textwidth]{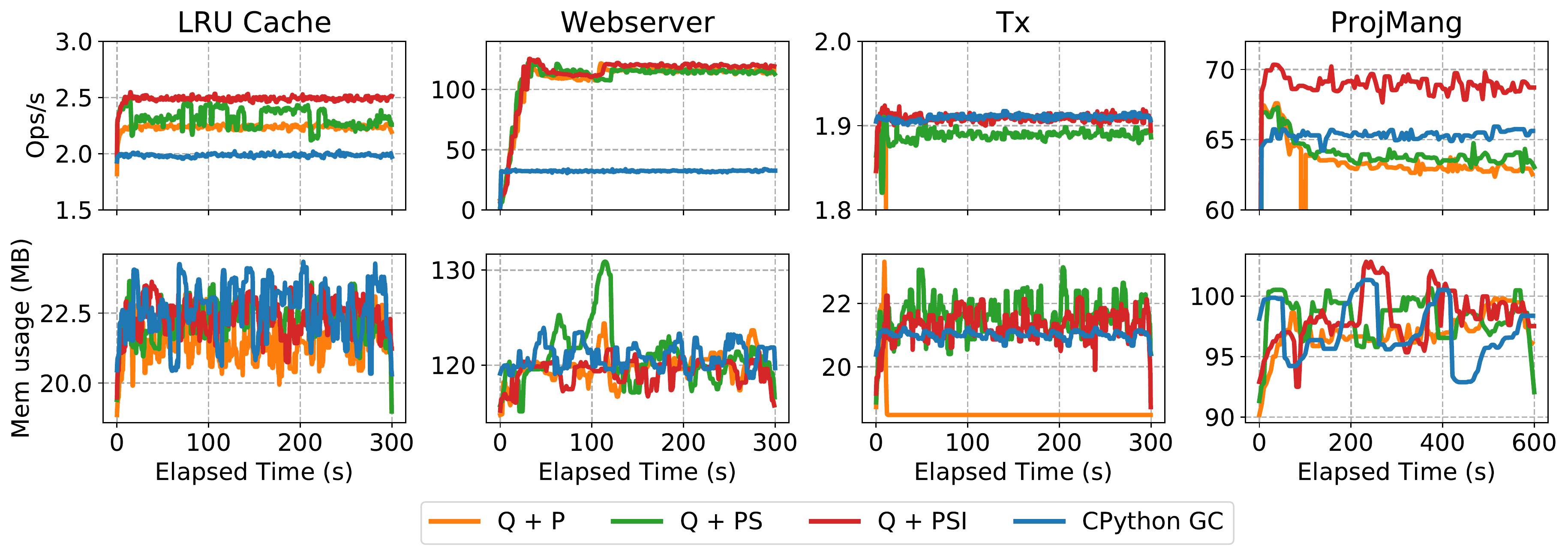}
  \caption{Reward and memory performance of each variant (minus ``Q'') and the CPython garbage collector.}
  \label{fig:gc_time}
\end{figure*}

We implemented a prototype learned garbage collector on top of CPython 3.7.5. Here, we present a preliminary experimental study, demonstrating that our implementation can learn a policy competitive with the built-in CPython GC for certain programs. We note that while the CPython GC is fairly simple, and may not represent a state-of-the-art garbage collector, the CPython GC is nevertheless an interesting baseline due to its widespread use. We use a learning rate of $\alpha = 0.1$ and a discount factor of $\gamma = 0.9999$.

\sparagraph{Test programs} We used three synthetic benchmarks and one real application for testing. Rewards are reported every two seconds. The memory threshold $M$ was set to the observed median memory usage under the CPython GC.

\begin{itemize}
\item{\textbf{LRU Cache} (synthetic): maintains large objects (containing cyclic references) in an LRU cache. Random queries, including lookups and inserts, are generated. The reward is  queries per second.}
\item{\textbf{Webserver} (synthetic): a simple HTTP server that responds to requests with a small webpage. This webpage is requested repeatedly by a number of clients. The reward is requests per second.}
\item{\textbf{Tx} (synthetic): searches for cycles in randomly generated transaction graphs. The reward is transactions searched per second.}
\item{\textbf{ProjMang} (real): a full scale, open source project management application.\footnote{\url{https://taiga.io/}} A workload of requests is generated and executed with a pool of concurrent workers. The reward is requests per second.}
\end{itemize}

\sparagraph{Variants} We tested four variants of our learned garbage collector. ``Q'' represents textbook Q learning with none of the optimizations described in Section~\ref{sec:optimizations}. ``Q + P'' represents Q learning with the ``prior'' optimization. ``Q + PS'' represents Q learning with the ``prior'' and ``reward shaping'' optimizations, and ``Q + PSI'' represents Q learning with all three of the optimizations described in Section~\ref{sec:optimizations}.
%\tim{How state of the art is Python's GC? Probably ok for a short paper, but otherwise, I would have liked to see some alternative baselines}

\sparagraph{Experimental setup} We ran each program with each GC variant for 5 minutes (except for ProjMang, which we ran for 10 minutes to achieve consistent results) and tracked the user-provided reward function and the memory usage of each learned GC variant and the CPython garbage collector. Programs were executed on a \texttt{c2-standard-16} virtual machine via the Google Cloud Platform.~\footnote{\url{https://cloud.google.com}}

\begin{table}
  \centering
  \begin{tabular}{lcccc}
    \toprule
    Application & Q & Q + P & Q + PS & Q + PSI \\ 
    \midrule
    LRU Cache & \bad{-94.96\%} & \good{12.86\%} & \good{17.34\%} & \good{25.48\%}  \\
    Webserver & \bad{-99.99\%} & \good{256.1\%} & \good{254.3\%} & \good{268.2\%} \\
    Tx        & \bad{-82.01\%} & \bad{-28.49\%}  & \bad{-1.12\%}  & \bad{-0.01\%} \\
    ProjMang  & \bad{-94.84\%} & \bad{-3.43\%}  & \bad{-2.22\%}  & \good{5.42\%} \\
  \end{tabular}
  \caption{Median improvement in reward function compared to the CPython garbage collector.}
  \label{tab:median}
\end{table}

\sparagraph{Median behavior} Table~\ref{tab:median} shows the median percent improvement in the reward function each variant achieved compared to the CPython GC over the entire runtime of the program. For example, the ``Q + PSI'' variant lead to around a 5\% increase in requests per second for the ProjMang benchmark. Unsurprisingly, the unoptimzied ``Q'' variant is unable to outperform the CPython GC on any benchmark, and suffers from vastly degraded performance. Of each of the four variants, ``Q + PSI'' consistently has the best performance. 

The most drastic improvement is seen on the Webserver benchmark. Here, the learned GC converges to a policy that does no garbage collection between when a request is received and when the response is queued for delivery by the operating system. Garbage collection is performed immediately after this point, effectively overlapping I/O with the GC computation. We note that this policy was fully learned -- the algorithm had no prior knowledge of I/O, system calls, or sockets. We also note that our experiment may be oversimplistic, as we used a simple webserver library\footnote{\url{https://flask.palletsprojects.com/}} that may not fully take advantage of Python's asynchronous I/O APIs.

No variant was able to improve on the CPython GC on the Tx benchmark (although some variants result in only a minor slowdown). By design the Tx benchmark generates a large number of long-lived objects containing cyclic references, which causes most collections to take a long time. This causes explorations made by the learned GC to be especially costly.

\sparagraph{Behavior over time} Figure~\ref{fig:gc_time} shows the reward and memory usage behavior of each variant (omitting ``Q'', which performed too poorly to be graphed on the same axes as the other variants) and the CPython GC. While many reinforcement learning techniques can require hours or days to train~\cite{neo,lift,decima}, each optimized learned GC variant is able to learn a competitive policy quickly, often within seconds.

The second row of Figure~\ref{fig:gc_time} shows memory usage over time, tracked via the CPython heap. Learned variants use slightly more memory (2\% on average) than CPython's GC, but this is after taking into account the size of the Q table and action buffer (4 - 18MB in our experiments). Future work focusing on compressing this table may be able to achieve improved performance \emph{and} lower memory usage.

\sparagraph[?]{Why not just tune the CPython GC} We found the default tuning parameters of the CPython GC to be close to optimal. After an exhaustive grid search, we were unable to find any tuning of CPython GC that improved performance without spacing out collections so much that memory usage ballooned. It is unsurprising that the default settings are performant, as they have been optimized for general use.

%%% Local Variables:
%%% mode: latex
%%% TeX-master: "main"
%%% End:

\vspace{-2mm}
\section{Future Work and Conclusion}
\label{sec:conclusions}

We have presented a initial prototype of a garbage collector powered by reinforcement learning. Our design incorporates a user-defined reward function, and customizes itself to the user's application on the fly using tabular Q learning.

Many challenges remain. In the future, we plan to further optimize the training and inference process described here. To tackle larger programs, more advanced methods, such as overlapping table updates with program execution, may be required. Further, we plan to investigate how more advanced reinforcement learning algorithms could be effectively applied to garbage collection. Finally, we hope to greatly expand our experimental analysis to a wider ranger of real-world applications, and further study the policies that are discovered by the learned garbage collector.

%%% Local Variables:
%%% mode: latex
%%% TeX-master: "main"
%%% End:

%\section*{Acknowledgments}
%
%TODO: Anybody to ack?

\balance
\clearpage

\bibliographystyle{abbrvnat}
\bibliography{ryan-cites-long}

\end{document}